\documentclass[pre,showpacs,twocolumn,rotating]{revtex4}
\usepackage{dcolumn}
\usepackage{graphicx}
\usepackage{amsmath}

\newcommand{\opr}[1]{\mathaccent 94 #1}

\newcommand{\ket}[1]{\left|#1\right\rangle}
\newcommand{\bra}[1]{\left\langle#1\right|}

\newcommand{\avr}[1]{\left\langle #1\right\rangle}

\newcommand{\D}{\displaystyle}
\begin{document}
\sloppy
\title{Simulation of wavepacket tunneling of interacting identical particles}
\author{Yu.E.Lozovik$^1$}
\email{lozovik@isan.troitsk.ru}
\author{A.V.Filinov$^{1,2}$}
\email{alex@ravel.mpg.uni-rostock.de}
\author{A.S.Arkhipov$^1$}
\email{antoncom@id.ru}
\affiliation{$^1$ Institute of Spectroscopy RAS, Moscow region, Troitsk, Russia, 142190}
\affiliation{$^2$ Fachbereich Physik, Universit$\ddot a$t Rostock Universit$\ddot a$tsplatz 3, D-18051 Rostock, Germany}
\date{\today}
\begin{abstract}
We demonstrate a new method of simulation of nonstationary quantum
processes, considering the tunneling
of two {\it interacting identical particles}, represented by wave
packets. The used method of quantum molecular dynamics (WMD) is based on
the Wigner representation of quantum mechanics. In the context
of this method ensembles of classical trajectories are used to solve
quantum Wigner-Liouville equation. These classical trajectories obey
Hamilton-like equations, where the effective potential consists of the usual classical
term and the quantum term, which depends on the Wigner function and its
derivatives. The quantum term is calculated using local distribution
of trajectories in phase space, therefore classical trajectories
are not independent, contrary to classical molecular dynamics.
The developed WMD method takes into account the influence of exchange and
interaction between particles. The role of direct and exchange interactions in
tunneling is analyzed. The tunneling times for interacting particles are calculated.
\end{abstract}
\maketitle
\section{Introduction}\label{secIntroduction}
A new quantum molecular dynamics method (QMD) was recently
successfully applied to a single wavepacket tunneling
\cite{DonosoMartens2001,LozovikFilinov1999}. This method is based
on the Wigner representation \cite{Lee1995,Wigner1932} of quantum
mechanics (further refered to as WMD - the Wigner representation
based MD). In the present paper we further develop this method and
consider its application to the {\it many-body} problem of
nonstationary tunneling of interacting identical particles.
Nonstationary tunneling is a problem of great interest in
particular in connection with developments of nanoelectronics.
Until now role of interaction and exchange in nonstationary
tunneling is not clear. To clear up this question is one the aims
of this paper. In this connection we consider the tunneling of two
identical charged particles, represented by wavepackets.

In the Wigner representation of quantum mechanics the state of the
system is described by the Wigner function, which obeys
Wigner-Liouville equation. The equation can be rewritten in the
form analogous to classical Liouville equation for classical
distribution function. This analogy is the basis of
WMD: the ensembles of classical trajectories are used to solve numerically
quantum Wigner-Liouville equation. The trajectories can be determined by equations
of motion analogous to classical ones. The used modification as against
classical equations of motion for the trajectories is an addition of extra
`quantum' term in the expression for the force \cite{DonosoMartens2001}.
This `quantum' term is expressed through the local approximation of the Wigner
function. For the approximation of the Wigner function we used
multi-dimensional Gauss distribution with the parameters determined through
the local moments of the ensemble of classical trajectories.

In the present paper the wavepackets moving in double-well
potential were considered. The interparticle interactions are
fully taken into account. The wavepackets are initially placed in
the same well on the one side of the barrier. We analyzed the
long-time evolution of wavepackets (for time scales corresponding
to many periods oscillation in the well) and consider the
probability to detect a particle in the first and in the second
well, respectively. Besides we study the short-time evolution
(characteristic times of interaction of wavepacket with the
barrier) and regard tunneling times.

Tunneling time is one of the most important features of nonstationary tunneling.
However, the theoretical definition of this quantity is nontrivial. There are
exist a lot of definitions of tunneling time
\cite{LM1992,SokConnor1993,Baz'1967,Rybachenko1967,But1982,Buttiker1983,HaugeStovn1989,
ButLan1986,AharonovBohm1961,Wlodarz2002,BEMuga2002,LLBR2002}. We use two common
approaches to determine tunneling time, namely {\em presence} and {\em arrival} times
(see \cite{DelgadoMuga1997,LozovikFilinov1999} and references therein).

First, one can consider the detector which reacts to the presence
of particles at some point $x_0$. The values measured by this
detector in a set of experiments on, e.g., particles transmission
through a barrier, would depend on particle density $\rho(x_0,t)$
at time $t$ and the mean {\em presence time} of a particle at
point $x_0$ would be given by
\begin{equation}
\langle t_p(x_0)\rangle = \D\int\limits_0^\infty dt\, t\, \rho(x_0,t) {\Big/}
\D\int\limits_0^\infty dt\, \rho(x_0,t)
\label{SitTime}
\end{equation}
For two points $x_a$ and $x_b$ one can consider the average time of transmission:
\begin{equation}
\langle t_T(x_a,x_b)\rangle = \langle t_p(x_b)\rangle-\langle t_p(x_a)\rangle
\label{TransTime}
\end{equation}
If the points are located on the different sides of a barrier,
then expression~(\ref{TransTime}) is an approximation for tunneling time.

Second, the detector that measures flux density can be used. For
this set of experiments one need to define another quantity - {\em
arrival time}. In this case the flux density operator must be
considered
\begin{eqnarray}
\opr J(x_0) = \frac 1 2[\opr p\delta(\opr q - x_0)+\delta(\opr q - x_0)
\opr p]\label{OprJ}\\
J(x_0,t) = \frac{\langle \psi(t)|\opr J(x_0)|\psi(t)\rangle}
{\D\int\limits_0^\infty dt \, \langle \psi(t)|\opr J(x_0)|\psi(t)\rangle},
\label{ArrivalTimeDistrib}
\end{eqnarray}
and the {\em arrival time} at point $x_0$ can be defined as
\begin{equation}
\langle t_a(x_0)\rangle= \int\limits_0^\infty dt\, t\, J(x_0,t) {\Big/}
\D\int\limits_0^\infty dt\, J(x_0,t)
\end{equation}
We stress that $J(x_0,t)$ can be negative due to the opposite
flux. Therefore Eq.~(\ref{ArrivalTimeDistrib}) can be used
directly as a probability distribution of {\em arrival times} only
if the opposite flux is negligible. This requirement can be
fulfilled if the detector is located far from the barrier. Then
Eq.~(\ref{ArrivalTimeDistrib}) determines the quasidistribution of
the {\em arrival times}. But in this case one can not distinguish
the time of transmission under the barrier and time of passing the
region between the barrier and the detector - still unresolved
problem of time measurement in quantum mechanics (see, e.g.,
\cite{HaugeStovn1989}). We use the {\em presence} and {\em
arrival} times from all variety of possible definitions of
tunneling time because their measurement in the framework of WMD
is relatively simple and, what is more important, the physical
meaning of the Eq.~(\ref{SitTime}) and~(\ref{ArrivalTimeDistrib})
is transparent and connected with the use of point-like detectors
in the set of the experiments on particles transmission.

By changing the strength of interaction between the particles, we
investigate the influence of interaction on tunneling. We also
consider the role of exchange. We found that the exchange is
important if the interaction is weak. In this case exchange has a
substantial influence on both the tunneling probability and
tunneling time. With the increase of interaction initial system
energy with fixed initial wave functions becomes greater. This
leads to decrease of tunneling times, the role of exchange gets
smaller and tunneling becomes insignificant in comparison with the
passing above the barrier. Our investigation had shown that WMD is
advantageous method, which can be used to solve the many-body
problems without enormous computer resources, and  which allows to
take into account such essentially quantum features as exchange
and tunneling.

We present here the investigation of the two-particle problem, but
the generalization of WMD for the case of more particles is
straightforward. The advantage of using the Wigner representation
in comparison with direct numerical solution of Schr\"odinger
equation is the following. Using WMD one does not need to store
large number arrays as with the grid methods. The basic algorithm
of WMD is very close to that of the common molecular dynamics
(MD), the distinction is only in the calculation of the force and
in the probability interpretation of initial conditions. During
about 40 years the classical MD methods were sufficiently improved
and all advantageous numerical schemes can be simply implemented
in WMD. The modern MD techniques allows to operate with the
thousands of particles and the same can be in principle achieved
by means of WMD, but in the last case one can consider quantum
particles.

We describe the simulation method and the physical model  in Sec.\ref{Method}
and Sec.\ref{Model}, respectively. Results are presented and
discussed in Sec.\ref{Results}. Main conclusions are summarized in Sec.\ref{secConclusion}.

\section{Simulation method}\label{Method}
\subsection{Equations of motion for Wigner trajectories}
The Wigner representation of quantum mechanics is one of the representations
which uses quantum distribution function in phase space. The Wigner
function $F^W(q,p,t)$ describes time evolution of the system and average
values of physical quantities are calculated with the help of scalar
functions, Weyl symbols $A^W(q,p)$:
\begin{equation}
\avr{A} = \int dp \int dq\, A^W(q,p)\, F^W(q,p,t)
\label{trRhoAW}
\end{equation}
It can be shown \cite{Lee1995,Tatarsky1983}, that Weyl symbols are
expressed through corresponding operators $A(\opr q, \opr p)$ as
follows:
\begin{equation}
A^W(q,p) = \frac{\hbar}{2\pi} \int d\xi \, d\eta \,
Tr\left[ A(\opr q,\opr p)\, e^{i\xi\opr q+i\eta\opr p}\right]\,
e^{-i\xi q-i\eta p}
\label{Weyl}
\end{equation}

The Wigner function is real and satisfies the following rules
\begin{eqnarray}
\int dp\, F^W(q,p,t) = \langle q|\opr\rho|q\rangle, \\
\int dq\, F^W(q,p,t) = \langle p|\opr\rho|p\rangle,
\label{Marg}
\end{eqnarray}
here $\opr\rho$ is the density operator. The Wigner function
$F^W(q,p,t)$ is also not nonnegative. There are nonnegative quantum distribution
functions, for example Husimi function \cite{Husimi1940}, but its evolution
equation is usually more complicated as against the Wigner function.

If one considers the Hamiltonian $H = p^2/(2m)+V(q)$, then the
evolution equation for the Wigner function (Wigner-Liouville equation) has the
form \cite{Lee1995,Tatarsky1983}:
\begin{equation}
\frac{\partial F^W}{\partial t}+\frac p m \frac{\partial F^W}
{\partial q} = \sum_{n=0}^\infty \frac{(\hbar/2i)^{2n}}{(2n+1)!}
\frac{\partial^{2n+1}V}{\partial q^{2n+1}}\frac{\partial^{2n+1}F^W}
{\partial p^{2n+1}}
\label{WigEvol3}
\end{equation}

If the potential does not have the terms with more than the second power of $q$,
then Eq.~(\ref{WigEvol3}) has the same form as for a classical distribution function $f$:
\begin{equation}
\frac{\partial f}{\partial t}+\frac p m \frac{\partial f}
{\partial q} = \frac{\partial V}{\partial q}\frac{\partial f}
{\partial p}
\label{ClasDistribEq}
\end{equation}
The Wigner function must satisfy a number of conditions
\cite{Tatarsky1983}, therefore initial function $F^W(q,p,t = 0)$
can not be chosen arbitrary. Even if $F^W(q,p,t)$ satisfies
`classical' equation (\ref{ClasDistribEq}) (for specific potential
$V$) it describes quantum system adequately because all quantum
corrections (all powers of $\hbar$) are held in the initial Wigner
function $F^W(q,p,t = 0)$. For example, the uncertainty principle
holds.

One can rewrite (\ref{WigEvol3}) in the form analogous to
(\ref{ClasDistribEq}):
\begin{equation}
\frac{\partial F^W}{\partial t}+\frac p m \frac{\partial F^W}
{\partial q} = \frac{\partial V_{eff}}{\partial q}\frac{\partial F^W}{\partial p},
\label{WigEvolVeff}
\end{equation}
where a new effective potential $V_{eff}$ is introduced
\begin{eqnarray}
\frac{\partial V_{eff}}{\partial q}\frac{\partial F^W}{\partial p} =
\frac{\partial V}{\partial q}\frac{\partial F^W}{\partial p}-
\frac{\hbar^2}{24}\frac{\partial^3 V}{\partial q^3}\frac{\partial^3 F^W}
{\partial p^3}+\cdots\label{DefVeff}
\end{eqnarray}

The characteristics of Eq.~(\ref{ClasDistribEq}) obey the equations
coinciding with classical equations of motion
\begin{equation}
\frac{\partial q}{\partial t} = \frac{p}{m}; \quad
\frac{\partial p}{\partial t} = -\frac{\partial V(q,p,t)}{\partial q}
\label{ClasMotion}
\end{equation}
From Eq.~(\ref{WigEvolVeff}) one can obtain the modified equations of motion for
Wigner trajectories \cite{Lee1995}
\begin{equation}
\frac{\partial q}{\partial t} = \frac{p}{m}; \quad
\frac{\partial p}{\partial t} = -\frac{\partial V_{eff}(q,p,t)}{\partial q}
\label{WigVeffMotion}
\end{equation}

To get information about time evolution of the system we
numerically solve equations (\ref{WigVeffMotion}) for the ensemble
of trajectories. To simplify our calculation of
$V_{eff}$~(\ref{DefVeff}) for the problem of interest we choose
the analytical form of the external potential and the interaction
between particles to contain only the 2-nd and the 4-th powers of
coordinates. In this case only the first two terms in the r.h.s.
of Eq.~(\ref{WigEvol3}) are non zero. As a result the total force
is a sum of the usual classical force and the `quantum' force
$F_k^{quant}$, which is infinite series in general, but in our
case contains only one term:
\begin{equation}
F_k^{quant} = \left(\frac{\hbar^2}{24}\right)\frac{\partial^3 V}{\partial q_i
\partial q_l\partial q_k}\frac{\partial^2 F^W}{\partial p_i\partial p_l}
\frac{1}{F^W},
\label{EfForce}
\end{equation}
where index $k$ is the $k$-th component of the force vector (there are $N\times d$ such components,
$N$ is the number of particles and $d$ is
spatial dimensionality), repetition of indexes indicates the summation.

As one can note the `quantum force' depends on the Wigner function, which is
unknown. To overcome this problem we use a local approximation for the Wigner
function in the vicinity of phase space point $x_a$ by Gaussian \cite{DonosoMartens2001}:
\begin{equation}
F^W(q,p,t) = F^W_0\,e^{-[(x-x_a(t))A_a(t)(x-x_a(t))+b_a(t)(x-x_a(t))]},
\label{WigLocApproxim}
\end{equation}
where $x = {q\choose p}$ is vector of all particle coordinates and
momenta, matrix $A_a$ (in our case of dimensionality $4\times 4$)
and vector $b_a$ (with dimensions $4\times1$) are obtained from
the local moments of the ensemble of trajectories in the vicinity of point
$x_a$.

\subsection{Consideration of exchange}
Exchange effects in this method can be in some cases considered simply by
using special initial conditions. Consider the system with the wave function $\Psi(x,t)$.
One can obtain the Wigner function as \cite{Lee1995}
\begin{equation}
F^W(q,p,t) = \frac{1}{(2\pi\hbar)^N}\int d\xi \, e^{ip\xi/\hbar}\,\Psi^*(q+\frac \xi
2,t)\,\Psi(q-\frac \xi 2,t)
\label{Wf}
\end{equation}
If the system consists of either bosons or fermions, wave functions
must be symmetrical or anti-symmetrical. If we regard the case when the Hamiltonian
does not depend on spins of the particles, then we can consider only the coordinate part of
wave function. Depending on the overall spin the coordinate part of wave function
is either anti-symmetrical or symmetrical. For example, wave function of the following
form is symmetrical (anti-symmetrical):
\begin{equation}
\ket{\Psi(1,2)} = \frac{\ket{\phi_1(1)}\ket{\phi_2(2)}\pm\ket{\phi_1(2)}\ket{\phi_2(1)}}
{\sqrt{2(1\pm|\bra{\phi_1}\phi_2\rangle|^2)}},
\label{SymWaveF}
\end{equation}
where $(i)$ means the dependence on variables of the i-th particle.
We use this wave function for the initial system state with
$\ket{\phi_k}$ of the form of a Gaussian wavepacket. As a result the Wigner function
takes the form:
\begin{widetext}
\begin{eqnarray}
F^W(q_1,q_2,p_1,p_2) =
\frac{1}{2(1\pm|\bra{\phi_1}\phi_2\rangle|^2)(2\pi\hbar)^2}\int dx_1 \,dx_2\,
e^{\frac i \hbar (p_1x_1+p_2x_2)}\,
\left[ \phi_1^*\left(q_1+\frac{x_1}{2}\right)\times
\right.\nonumber\\
\left.\phi_2^*\left(q_2+\frac{x_2}{2}\right)\pm\phi_1^*\left(q_2+ \frac{x_2}{2}\right)
\phi_2^*\left(q_1+\frac{x_1}{2}\right)\right]
\left[\phi_1\left(q_1-\frac{x_1}{2}\right)
\phi_2\left(q_2-\frac{x_2}{2}\right)\pm \phi_1\left(q_2-\frac{x_2}{2}\right)
\phi_2\left(q_1-\frac{x_1}{2}\right)\right]
\label{WFIntPsi12}
\end{eqnarray}
and can be rewritten as
\begin{eqnarray}
F^W(q_1,q_2,p_1,p_2) =
\frac{1}{2(1\pm|\bra{\phi_1}\phi_2\rangle|^2)}
\left[W_1(q_1,p_1)\,W_2(q_2,p_2)+\right.\nonumber\\
\left.W_1(q_2,p_2)\,W_2(q_1,p_1)\pm U_{12}(q_2,p_2)\,U_{21}(q_1,p_1)\pm U_{12}(q_1,p_1)\,
U_{21}(q_2,p_2)\right],
\label{WFWiUij}
\end{eqnarray}
\end{widetext}
where
\begin{eqnarray}
W_k(q,p) = \frac{1}{(2\pi\hbar)}\int d\xi \, e^{ip\xi/\hbar}\, \phi_k^*(q+\frac \xi
2)\, \phi_k(q-\frac \xi 2)\label{WiF}
\end{eqnarray}
and
\begin{eqnarray}
U_{kj}(q,p) = \frac{1}{(2\pi\hbar)}\int dx\, e^{ipx/\hbar}\, \phi_k^*(q+\frac x
2)\, \phi_j(q-\frac x 2)
\label{UijF}
\end{eqnarray}

In coordinate space the initial state (\ref{SymWaveF}) is described by
wave function of the following form:
\begin{eqnarray}
\phi_k(x) = \frac{1}{(2\pi\hbar)}exp\left(-\frac{(x-x_{k0})^2}{4\sigma_k^2}+\frac{ip_{k0}
(x-x_{k0})}{\hbar}\right)\label{GaussWaveF}
\end{eqnarray}
For this case,
\begin{eqnarray}
W_k(q,p) = \frac{1}{\pi\hbar}exp\left(-\frac{(q-x_{k0})^2}
{2\sigma_k^2}-\frac{(p-p_{k0})^2}{2(\hbar/(2\sigma_k))^2}\right)
\end{eqnarray}
and the term with $U_{kj}$ in Eq.~(\ref{WFWiUij}) is proportional to
$exp\left(-A(x_{10}-x_{20})^2\right)$,
where $A$ is a positive constant. If $\sigma_1 = \sigma_2 = \sigma$, then
$A = 1/(2\sigma^2)$. For $|x_{10}-x_{20}| >> \sigma$ this term can be
neglected and one gets:
\begin{eqnarray}
F^W(q_1,q_2,p_1,p_2) = \frac{1}{2(1\pm|\bra{\phi_1}\phi_2\rangle|^2)}\times\nonumber\\
\left[W_1(q_1,p_1)\,W_2(q_2,p_2)+W_1(q_2,p_2)\,W_2(q_1,p_1)\right]
\label{WFGaussTwoP}
\end{eqnarray}
We emphasize that this approximation is used only at the initial time
moment. Further the dynamical equations are solved formally
exactly.

In the considered problem two particles move in the potential
\begin{eqnarray}
U(x) = \alpha(-x^2+\gamma x^4), \quad \alpha, \gamma > 0
\end{eqnarray}
The potential of interparticle interaction is: \\
$V_{int}=\{const - \beta r^2\}$, if $\{const - \beta r^2\}>0$, and $=0$, otherwise.
If we disregard discontinuity in the interparticle potential then the
distinction from harmonic oscillator is the 4-th power of $x$ and one has only one quantum term
in the force (\ref{EfForce}). Using the classical trajectories and the Gaussian approximation for
the Wigner function one can solve the Wigner-Liouville equation `exactly'. The distinction
of the adopted approximation from the case of distinguishable particles is that now initial
positions of two particles may be in the Gaussian centered at $x_{10}$ or at $x_{20}$. In this
way we regard the symmetry in exchange of particles and obtain the picture of their motion.

\subsection{Algorithm and calculation of average values}
Our simulation algorithm is the following. First, the initial coordinates and momenta of every
trajectory in the ensemble are distributed according to chosen parameters of
the wavepackets (mean coordinate, momentum and their variances). Second, we
calculate the `quantum force' and solve numerically equations of motion.

For the $j$-th trajectory at time $t$ with coordinates and momenta
$\{q^{(j)}(t)$, $p^{(j)}(t)\}$ one has to compute the local moments of the ensemble
of trajectories in the vicinity of the point $\{q^{(j)}(t)$, $p^{(j)}(t)\}$
(point $x_a$), using the weight function which rapidly goes to zero with the
increase of distance to this point (uncertainty principle must hold) \cite{DonosoMartens2001}.
The approximation~(\ref{WigLocApproxim}) is the many-dimensional Gauss distribution
of the vector $x = {q\choose p}$. Matrix $A_a = \frac 1 2 C_a^{-1}$,
$C_a^{-1}$ is the inverse matrix of covariance $C_a$, and vector
$b_a = -2A_af_a$, where $f_a$ is the vector of averages. If one calculates
$\langle q_i\rangle$, $\langle p_i\rangle$, $\langle q_iq_k\rangle$,
$\langle p_ip_k\rangle$, $\langle q_ip_k\rangle$, $i,k = 1,\dots Nd$,
one obtains $f_a = \langle{q-q_a\choose p-p_a}\rangle$ and symmetrical matrix
$C_a$ with elements
\begin{eqnarray}
C_a(l,m) = \langle (q_l-q^{(a)}_l)(q_m-q^{(a)}_m)\rangle -\nonumber\\
\langle q_l-q^{(a)}_l\rangle\langle q_m-q^{(a)}_m\rangle
\end{eqnarray}
for $l = 1,\dots Nd, \; m = 1,\dots l$,
\begin{eqnarray}
C_a(l,m) = \langle (p_{l-Nd}-p^{(a)}_{l-Nd})(p_{m-Nd}-
p^{(a)}_{m-Nd})\rangle -\nonumber\\
\langle(p_{l-Nd}-p^{(a)}_{l-Nd})\rangle\langle(p_{m-Nd}
-p^{(a)}_{m-Nd})\rangle
\end{eqnarray}
for $l = Nd+1,\dots 2Nd,\; m = Nd+1,\dots l$, and
\begin{eqnarray}
C_a(l,m) = \langle (q_l-q^{(a)}_l)(p_{m-Nd}-p^{(a)}_{m-Nd})
\rangle-\nonumber\\
\langle (q_l-q^{(a)}_l)\rangle\langle(p_{m-Nd}-
p^{(a)}_{m-Nd})\rangle
\end{eqnarray}
for $l = 1,\dots Nd, \; m = Nd+1,\dots 2Nd$. Here $q_i$ and $p_i$ are
the $i$-th components of vectors of all coordinates $q$ and momenta $p$;
$\langle \dots \rangle$ means the averaging over all trajectories with the
weight function, which rapidly approaches zero with growing distance to $x_a$ in phase
space. After that, one can calculate the inverse matrix for $C_a$ and get $A_a$
and $b_a$.

At every time $t$ for the $j$-th trajectory with coordinates $q^{(j)}(t)$ and
momenta $p^{(j)}(t)$ matrix $A^{(j)}(t)$ and vector $b^{(j)}(t)$ are calculated, therefore
the `quantum force' for the $j$-th trajectory is known. Further, one has to solve
equations~(\ref{WigVeffMotion}), for example by Runge-Kutt and Adams methods.

To calculate average values we use the following approximation for the Wigner function:
\begin{equation}
F^W(q,p,t) = \frac 1 K \sum_{k=1}^K \delta\{q-q_k(t)\}\,
\delta\{p-p_k(t)\}
\label{WigAvApproxim}
\end{equation}
The summation is over all trajectories in the ensemble. One of the interesting values
characterizing tunneling is the reaction probability:
\begin{equation}
R(q_a, t)= \frac{1}{N}\int\limits_{q_a}^{\infty}\rho(x,t)\,dx, \label{ReactProb}
\end{equation}
where the lower limit of integration is the point of the largest height of the barrier,
$\rho(x,t)$ is the particle density at point $x$ and $N$ is the particle number. This
quantity shows what part of wavepackets are currently in the right well.

Another important value is {\em tunneling time}. In this paper to
determine {\em tunneling time} we use two methods: {\em presence}
and {\em arrival} times. In the Wigner formalism these quantities
are expressed through the following integrals (see~(\ref{SitTime})
and~(\ref{ArrivalTimeDistrib})):
\begin{equation}
|\psi(x_0,t)|^2 = \int dp \, F^W(x_0,p,t)\label{ProbDensWig}
\end{equation}
and
\begin{equation}
J(x_0,t) = \frac{\D\int dp\, p\, F^W(x_0,p,t)}
{\D\int\limits_0^\infty dt \D\int dp \, p\, F^W(x_0,p,t)}\label{ATDistribWig}
\end{equation}
(the Weyl symbol for flux operator has the form\\
$J_{x_0}=\frac{\hbar}{2}\sin\left(\frac{2p( x_0-q)}{\hbar}\right)\frac\partial{\partial q}\delta(q-x_0)$).

\section{Model problem}\label{Model}
\subsection{Hamiltonian}
We consider the following model problem. Two particles move in one-dimensional space ({\it e.g.}
in a quantum wire). The Hamiltonian of the system reads:
\begin{equation}
H = \sum_{i=1}^2 \left(\frac{p_i^2}{2m}+\alpha(-q_i^2 + \gamma q_i^4)\right)+
U\left(|q_1-q_2|\right),
\label{Ham1}
\end{equation}
where $q_1, q_2, p_1, p_2$ are particle coordinates and momenta, $U$
is interaction energy.
We use the system of units with $\{\hbar=m=\alpha=1\}$,
$l_0= \hbar^{1/2}/ (m \, \alpha)^{1/4}$ is the unit of length,
$E_0= \hbar(\alpha/m)^{1/2}$ is the unit of energy, and the unit
of time is $t_0 = (m/\alpha)^{1/2}$.

Initially particles are placed in the left well, their wave
functions have the form of Gaussian wavepackets. Initial mean
momenta and coordinates of wavepackets and their variance in
momentum and coordinate spaces are chosen to make the transmission
above the barrier as lower as possible and the overlapping of the
wavepackets is negligible. Particles move in the direction of the
barrier. This model roughly describes nonstationary tunneling of
two electrons through the potential barrier in a quantum wire or
tunneling between two quantum wells. This can be realized, for
example, when with the help of laser pulses one prepares the state
of two electrons in the form of two wave packets in nanostructure
and study the system evolution in time.

In the used system of units the Hamiltonian is:
\begin{equation}
H = \sum_{i=1}^2 \left(\frac{p_i^2}{2}-q_i^2 + \gamma q_i^4\right)+U(|q_1-q_2|)\label{Ham2}
\end{equation}

The Coulomb potential $Q_1Q_2/r$ describes the interaction between particles.
The problem becomes one-dimensional if characteristic energies of
the transverse quantization
are much larger
than the energies of the longitudinal motion. If it is valid then adiabatic approximation applies
and the problem is really one-dimensional. The interparticle interaction
then reduces to
\begin{eqnarray}
U(r) = \lambda\frac 2 {a^2}\int\limits_0^\infty\frac{e^{-\frac{\rho^2}
{a^2}}\rho}{\sqrt{r^2+\rho^2}}d\rho =\nonumber\\
\lambda\frac{\sqrt\pi} a e^{\frac{r^2}{a^2}}(1-erf(\frac r a)),
\label{Interact}
\end{eqnarray}
where we performed integration over the particle wavefunctions of transverse quantization,
with $a$ being the characteristic width of the quantum wire. The interaction parameter,
$\lambda = Q_1Q_2m^{3/4}/(\hbar^{3/2}\alpha^{1/4})$, is the ratio of characteristic Coulomb
energy and energy of oscillator.

In our model problem we substitute the potential (\ref{Interact}) for model quadratic one,
just to avoid uncertainties related with the calculation of the quantum force (\ref{EfForce})
and demonstrate the method for the case of exchange. Parameters of this potential are chosen
to make it as close as possible to expression~(\ref{Interact}):
\begin{equation}
U(r)=
\begin{cases}
\frac{\lambda}{a}(\sqrt\pi-0.05(r/a)^2),  & \text{if $r < a (20\sqrt{\pi})^{1/2}$}, \\
0, & \text{if $r \ge a (20\sqrt{\pi})^{1/2}$}.
\end{cases}
\end{equation}

\subsection{Initial parameters}
We analyze the system described by the  Hamiltonian (\ref{Ham2})
for two cases - with and without exchange, respectively. The main
quantity analyzed is the reaction probability (\ref{ReactProb}).
Its largest value is unity when both particles are entirely in the
right well. The reaction probability clearly shows the
`distribution' of particles between two wells and characterize
their time evolution.

Consider two cases, we call them {\em $0$-th order WMD} (Wigner
molecular dynamics in zeroth approximation) and {\em $n$-th order
WMD} (further simply refered to as the `{\em quantum case}'). For
{\em $0$-th order WMD} the quantum term in force (\ref{EfForce})
is neglected and evolution of the trajectories is determined by
classical Hamilton equations. Therefore only passing above the
barrier is taken into account. This approximation is not purely
classical, because the initial distribution is the same as for the
`{\em quantum case}': $|F^W(q,p,t=0)|$ can contain arbitrary
powers of $\hbar$. As a result we have a classical evolution of
the quantum distribution function, therefore we call this case
{\em $0$-th order WMD}, not the `classical' MD. The difference
between {\em $0$-th order} and {\em $n$-th order} is that in the
latter case the quantum term in the force is regarded.

The initial distribution in coordinate space has the form of two
Gaussian wavepackets, which practically do not overlap. In the
`{\em quantum case}' one can consider two situations. First, we
can consider the problem neglecting exchange ({\it i.e.} regard
the distinguishable particles). Second, we can take exchange into
account. In the first case the initial wave function is a product
of one-particle wave functions, and the Wigner function has the
form $F^W(q_1,q_2,p_1,p_2) = W_1(q_1,p_1)\, W_2(q_2,p_2)$ (compare
with Eq.~(\ref{WFGaussTwoP})). This means that one of the
Gaussians corresponds to the first particle and another to the
second one. For such initial distribution function exchange
effects can not arise and we will call this situation the `{\em
quantum case}' without exchange.

In the second situation the particles are identical and the
initial wave function is symmetrical (or antisymmetrical). Now the
Wigner function has the form (\ref{WFGaussTwoP}). Both Gaussians
may correspond either to the first or to the second particle. The
initial coordinates and momenta of some trajectory, $\{x_1, x_2,
y_1, y_2\}$, are chosen with the probability
$|F^W(q_1,q_2,p_1,p_2)|$ of the configuration $\{q_1=x_2, q_2=x_1,
p_1=y_2, p_2=y_1\}$. If the wavepackets are initially close to
each other, the terms $U_{kj}$ in Eq.~(\ref{WFWiUij}) do not
vanish and the procedure of setting the initial coordinates and
momenta becomes more complicated.

For both `{\em quantum cases}' (without and with exchange)
dynamical correlations are taken into account due to solution of
the Wigner-Liouville equation. Statistical correlations are not
regarded in the `{\em quantum case}' without exchange but they are
allowed in the case with exchange. In this sense two situations
resemble Hartree and Hartree-Fock approximations, respectively.
Note that we {\em do not use} mean-field approximation, the
similarity must be regarded only in the meaning formulated above.

For the {\em $0$-th order} WMD both ways of setting the initial
coordinates and momenta for trajectories can be applied,
but it was found that the result is practically independent on it.
We use the following parameters: initial coordinates and momenta for Gaussians
are $x_1(0)=-140, \,p_1(0)=45$ and $x_2(0)=-310,\, p_2(0)=90$,
dispersions in coordinate space ($\sigma_x$ for the Gauss function
$exp[-x^2/(2\sigma_x^2)]$) are the same for both wavepackets and equal $20$, in
momentum space $\sigma_p = \hbar/(2\sigma_x) = 0.025$.
Parameters of the external potential are $\alpha=1,\, \gamma=1.25\times 10^{-8}$.

One of our aims is to investigate the influence of interaction on
tunneling. We change the interaction parameter $\lambda$, starting
with $\lambda = 0$ (no interaction). Effective charge of electrons
or holes in nanostructures can be controlled by changing the
permittivity, but not in very wide range. Here we would like to
draw the attention to the fact that due to the use of special
system of units, $\hbar = m = \alpha = 1$, the region of variation
of the dimensionless interaction parameter can be pretty wide.
Actually, in this unit system the parameters of external potential
are used. Therefore we can vary the interaction $\lambda$ by
changing the external potential. This change leads to scaling of
the units of length, time, energy and so on.

\section{Results and discussion}\label{Results}
The definition of tunneling is usually given for one particle. We
analyze the motion of two particles and as the {\it under barrier
transmission} is of interest for us, the {\it total} initial
energy of the system is set about $0.99$ of the height of the
barrier. So even if one of particles borrows all the energy the
latter is still lower than the height of the barrier. We deal with
the wavepackets, therefore though the mean energy per particle is
lower than the barrier height, the transmission above the barrier
is still possible due to the dispersion in the momentum space. The
comparison between {\em $0$-th} and {\em $n$-th order} WMD allow
us to estimate the transmission under the barrier. The state of
the particles is considered as an entangled unified whole and the
'transmission under the barrier' means the transmission of at
least one of the particles.
\subsection{Reaction probabilities}
\begin{figure*}
 \includegraphics[height=10cm]{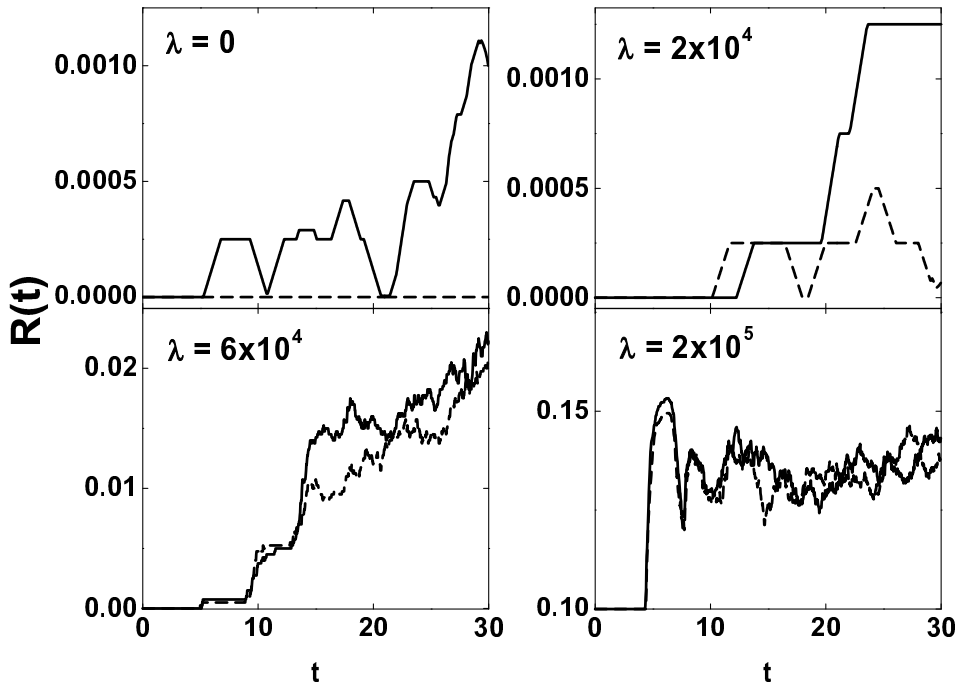}
 \caption{\label{Fig1}Time-dependence of the
 reaction probability, $R(t)$,  for the {\em $0$-th order} WMD
(dashed lines) and `{\em quantum case}' (solid lines). Interaction
 strengths are $\lambda = 0, \, 2\times 10^4,\, 6\times 10^4, \, 2\times 10^5$.
Maximum value of the reaction probability $R(t)=1$,
time $t$ is in the units $t_0 = (m/\alpha)^{1/2}$,
$\alpha$ is the potential parameter~(\ref{Ham1}).}
\end{figure*}

In Fig.~\ref{Fig1} we show the reaction probability for the
interaction $\lambda=0, \, 2\times 10^4,\, 6\times 10^4$ and
$2\times 10^5$. The `{\em quantum case}' (with exchange) and {\em
$0$-th order} WMD are compared. One can see that for weak
interaction there are a large difference between this two cases.
Under barrier transmission takes place only for the `{\em quantum
case}', in the {\em $0$-th order} only wavepacket components with
the energy above the height of the barrier can pass to the right
well.

With the increase of the interaction the reaction probability grows for both cases.
The reason is that the initial energy becomes
larger with the increase of $\lambda$ and
there are more wavepacket components with the energy above the barrier height.
For the `{\em quantum case}' it is also important that
there are some high-energy components which pass under the barrier.
For very large values $\lambda \geq 6\times 10^4$
reaction probabilities of the {\em $0$-th order} and the `{\em quantum case}' are
almost the same. It means that for strong interaction
the role of tunneling is negligible,
wavepackets include too many components, which can pass above the barrier.

\begin{figure*}
\includegraphics[height=10cm]{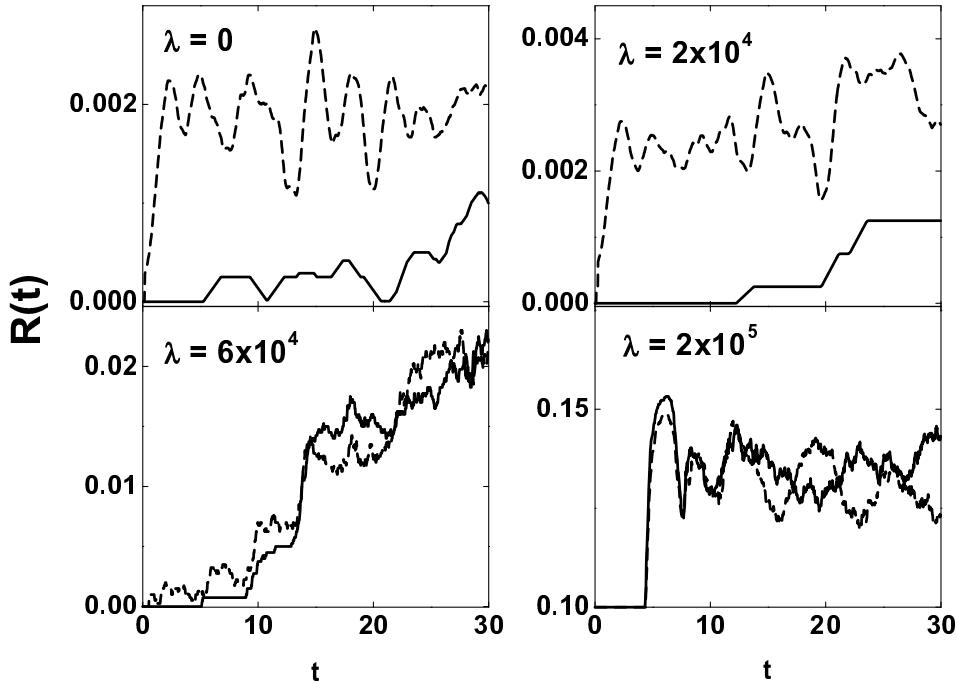}
\caption{\label{Fig2} Time-dependence of the reaction probability,
$R(t)$,  for the `{\em quantum case}' with exchange (solid lines)
and without exchange (dashed lines). Interaction strengths are
$\lambda = 0, \, 2\times 10^4, \, 6\times 10^4, \, 2\times 10^5$.}
\end{figure*}

In Fig.~\ref{Fig2} we present reaction probabilities for the `{\em quantum case}' with
and without exchange. For large $\lambda$ classical transmission above the barrier
prevails and the influence of exchange on tunneling is considerably small. For small $\lambda$
one can see a large difference in the reaction probabilities.
If the particles are distinguishable, the reaction probability is
larger, but the sign of the effect depends on the initial parameters
of the wavepackets, they can be fitted to make reaction probability
greater for the case with exchange. Initially the distribution of
the particles in coordinate space has the form of two separated
Gaussians. We found that, for the used parameters, these two peaks
quickly merge, forming a single peak, and very seldom they can
be seen as two separated wavepackets. It means that particles are
close to each other during most time of simulation and
exchange effects must be substantial. From Fig.~\ref{Fig2} one can see that
exchange almost does not influence the reaction probability for large
$\lambda$. In this situation the contribution of transmission
above the barrier to the reaction probability is very high (in comparison
with tunneling, compare with the {\em $0$-th order} WMD in Fig.~\ref{Fig1}).
So it is difficult to notice exchange effects against this background. It
is possible that the observed effect of the increase of
transmission above the barrier with the growth of the interaction parameter
masks the effect of the increase of dynamical correlations between
electrons, which also suppress exchange effects.

The reaction probabilities presented in Figs.~\ref{Fig1} and \ref{Fig2} were measured with the
finite precision, connected with the finite time step and finite number of the used trajectories.
Therefore the regions of constant reaction probability for small $\lambda$ arise - the subtle
noise was smoothed over the errors. The substantial features (differences) of {\em $0$-th order}
and {\em $n$-th order} WMD in cases {\em with} and {\em without exchange} become apparent on
much larger scales and they are easily taken into account. Probably, the observed noise is
due to the additional oscillations of high-energy parts of the wave packets.
It will be the subject of detailed investigations in the next work.

Another common feature of Figs.~\ref{Fig1} and \ref{Fig2} is the
behavior of the reaction probability for $\lambda = 2\times 10^5$.
At the end of considered time interval, $0 \leq t \leq 30$, the
reaction probabilities settle around the value $\approx 0.15$. It
seems that in the limit of infinite $\lambda$ the equilibrium
corresponds to the situation when both the wells are occupied with
the equal probability, because in this case the total initial
energy is much greater than the height of the barrier. Then the
reaction probability must settle around $0.5$. Probably, $\lambda
= 2\times  10^5$ is not large enough and the value of initial
total energy is sufficient only to push a small part of the wave packet through the barrier. It is possible that then the leakage to the
right well is balanced by the opposite transmission to the left
well and that is why the reaction probability, $R(t)$, stays near
the value $0.15$. On the other hand, there can be another
explanation: after the transmition of some part of the wavepackets
to the right well, the repulsion between this part and the part in
the left well prevents further penetration of particles to the
right well, as a result the reaction probability settles at
$\approx 0.15$ (i. e., mechanism analogous to Coulomb blockade).
Whether this second mechanism realizes is unclear, perhaps, these
two processes take place simultaneously. As for the first
scenario, the kinetic equilibrium, there are no doubts it can
exist: the reaction probability also settles around some value for
$\lambda = 0$ (the `{\em quantum case}' without exchange,
Fig.~\ref{Fig2}). For this case there is no interaction, therefore
the only explanation can be the equality of transmissions through
the barrier in both directions. In fact, there are no reasons to
believe that the reaction probability will always stay near, say,
$0.15$ (for $\lambda = 2\times  10^5$), possibly here we observe
only the intermediate equilibrium and later the system can come to
the state when the reaction probability is about $0.5$. But such
extra-long-time evolution must be the subject of special
investigation.

\subsection{Hartree description of the tunneling}
Both Fig.~\ref{Fig1} and Fig.~\ref{Fig2} demonstrate oscillations
of the reaction probability, with the growth of interaction
parameter $\lambda$ their period decreases and the picture becomes
less regular. This is due to the behavior of the transmitted part
of wavepackets: to the right from the barrier there is the wall of
the right well, transmitted part is reflected from it and moves to
the left. Then it is partially transmitted back to the left well,
making some modulation of the reaction probability curve.
Transmission takes place mainly when the wavepackets come to the
barrier, therefore reaction probability changes step by step with
some period. When the interaction is weak, the particles move
almost independently and this period coincides approximately with
that of oscillations of one particle in the well ($T\approx 4$,
see Figs.~\ref{Fig1} and \ref{Fig2}, the curves for $\lambda =
0$). But with the increase of $\lambda$ this period is changed
accordingly to the influence of one particle on the motion of the
other particle. This influence can be illustrated more
transparently in the picture of effective one-particle barriers
(see below).

\begin{figure*}
 \includegraphics[height=10cm]{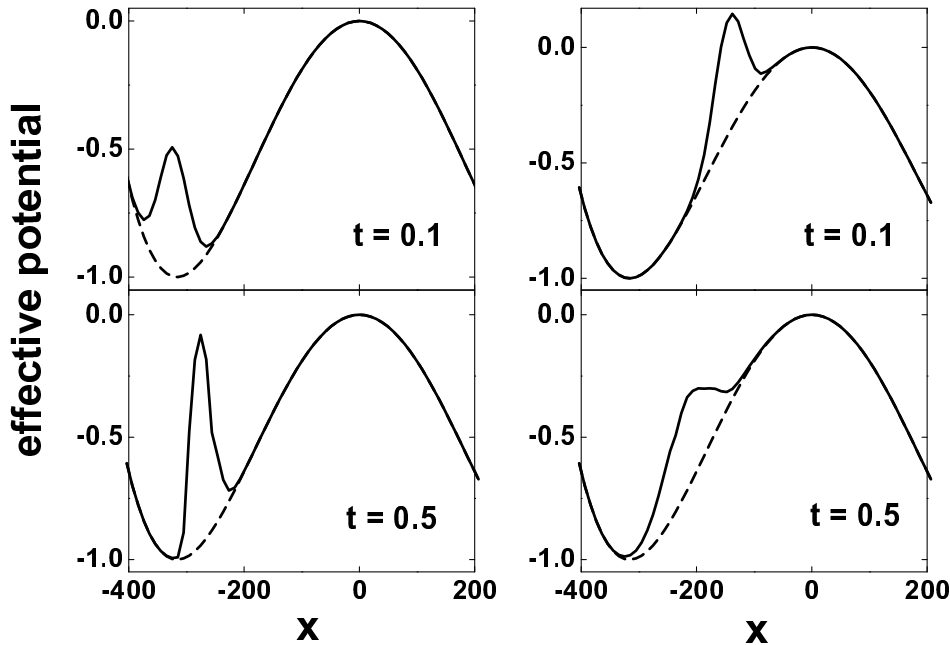}
 \caption{\label{Fig3} Effective potentials (\ref{Vef}) (solid
 lines), compared with the stationary external potential (dashed
 lines), in units of the height of the barrier. Two times,
 $t = 0$ and $t = 0.5$, are considered. Two left (right)
 plots are the effective potentials for the particle, which is initially closer
(farther) to (from) the  barrier. We consider the `{\em quantum case}'
without exchange, $\lambda = 2\times  10^5$, $x$ is in units $l_0 = \hbar^{1/2}
 (m\alpha)^{-1/4}$, $\alpha$ is the potential parameter~(\ref{Ham1}).}
\end{figure*}

In terms of Hartree approximation it means that each particle
moves effectively in the double-barrier potential. The first
barrier is the stationary barrier between the wells and the second
one is the effective time-dependent barrier due to the interaction
of particles. Each particle either fall on two barriers, or move
between them. In the latter case a particle is inside the
potential well. As the other particle moves closely, the well
becomes more narrow and the energy levels in it higher. Therefore
the energy of the particle in the well becomes greater and the
probability of transmission increases. Another mechanism is that
due to the nonadiabatic narrowing of the well the particle can
jump on the higher energy levels, which also results in the
increase of transmission probability.

The effective potential for the first particle in Hartree
approximation can be defined as
\begin{equation}
 V_{eff}(x_1,t)=V_{ext}(x_1)+\int U(x_1,x_2)|\, \psi(x_2,t)|^2 \, dx_2,
 \label{Vef}
\end{equation}
where $V_{ext}$ is the external potential, $U$ is the interaction
between particles, and $|\psi(x_2,t)|^2$ is the probability
density for the second particle. The effective potential for the
second electron is given by the analogous equation. Of course,
this can be applied only to the case when the particles are
distinguishable. If the particles are identical, one can use the
Hartree-Fock approximation, and the potential becomes nonlocal.
{\em We use neither Hartree nor Hartree-Fock approximation} in our
method, they are just very convenient tools of visualizing the
behavior of quantum particles, which interact with the barrier and
between each other.

In Fig.~\ref{Fig3} we plot the effective potentials (\ref{Vef}), $\lambda = 2\times  10^5$. The
dotted line is the shape of external potential. The `{\em quantum case}' without exchange is
considered, the particles are not identical, therefore {\em Hartree approximation} can be used. The
case with exchange is more interesting but if particles are identical, one can regard the
effective potential, which is the same for all particles. Here we just illustrate the possible
mechanisms of transmission and the case of distinguishable particles is more representative. In
this case one can differ two situations. First, the barrier, which arises due to
interparticle interaction, is close to the the stationary barrier
(effective broadening of the stationary barrier in the two right plots in Fig.~\ref{Fig3}).
Second, this effective barrier is
closer to the left wall of the well (the two left plots in Fig.~\ref{Fig3}). For identical
particles, these situations take place simultaneously and the effective potential is the same
for every particle.

The right plots in Fig.~\ref{Fig3} show that the interaction makes
the barrier wider and it prevents the transmission. But in the
left plots, the well becomes not so deep, the energy levels in it
grow and low-energy components of the wavepacket can be reflected from
the effective barrier in the direction of the stationary barrier.
The tunneling and transmission above the barrier from the higher energy
levels are stronger. The oscillations of the low-energy components
between the stationary and effective barriers make the tunneling
probability larger.

\subsection{Tunneling times}
Consider now Table~\ref{tab1} where we present tunneling times for the spatial interval $[-45, 45]$
determined by two methods: {\em presence}~(\ref{SitTime}) and {\em arrival}
times~(\ref{ArrivalTimeDistrib}). The interaction strengths are
$\lambda=0, \, 6\times 10^4, \, 2\times 10^5$. We consider
two situations: the `{\em quantum case}' with and without exchange.
At the edge points of this spatial
interval the value of the external potential approximately coincides with the initial energy
of the system. Two methods give close results, {\em tunneling time} calculated with the help of
Eq.~(\ref{ArrivalTimeDistrib}) is in general greater than that with use of Eq.~(\ref{SitTime}),
but the trend connected with changes in tunneling time with change of the
interaction parameter is the same for both methods. Tunneling time is greater
for the case with exchange, because for our parameters there are fewer
high-energy components in wavepackets and tunneling is weaker for this
case. With the increase of $\lambda$ the tunneling time gets smaller,
this can also be connected with the fact
that for strong interaction there are more high-energy components in the wavepackets. Those
components make the main contribution to the transmission
because the more the energy the greater the transmission probability.
Because the components with higher energy move
faster the time of passing some interval become smaller. We do not list in Table~\ref{tab1}
transmission times for the case of free motion. These times are all about $2.0$ and are
almost independent on $\lambda$, exchange and method of their calculation. One can see that the
presence of the barrier makes the transmission (tunneling) time smaller. It is due to enriching of
the transmitted part of the wavepacket by high-energy components, which has greater probability
to go through the barrier and move faster. Therefore on average transmitted part arrives to
detector {\it earlier} than the whole wavepacket in the case of free motion.

\begin{table}[th]
\caption{\label{tab1}Tunneling times for the spatial interval [-45,45]. (The
system of units is $\hbar = m = \alpha = 1$).\vspace*{1pt}}
{\footnotesize
\begin{tabular}{|c|c|c|c|}
\hline
{} &{} &{} &{}\\[-1.5ex]
{} & $\lambda=0$ & $\lambda=6\times 10^4$ & $\lambda=2\times 10^5$\\[1ex]
\hline
{} &{} &{} &{}\\[-1.5ex]
{presence time (no exchange)} &0.8($\pm$0.1) &0.72($\pm$0.09) &0.38($\pm$0.05)\\[1ex]
{arrival time (no exchange)} &0.9($\pm$0.1) &0.86($\pm$0.09) &0.80($\pm$0.08)\\[1ex]
\hline
{} &{} &{} &{}\\[-1.5ex]
{presence time (exchange)} &0.8($\pm$0.1) &0.78($\pm$0.08) &0.36($\pm$0.04)\\[1ex]
{arrival time (exchange)} &1.2($\pm$0.1) &1.06($\pm$0.09) &0.72($\pm$0.07)\\[1ex]
\hline
\end{tabular} }
\end{table}

\section{Conclusion}\label{secConclusion}
We analyzed the nonstationary tunneling of two interacting
identical particles by quantum molecular dynamics method based on the Wigner representation.
The WMD method allows to calculate different features characterizing quantum evolution
and influence of particle interaction and exchange on tunneling. We found that the strong
interaction in this problem leads to decrease of the role of tunneling in the transmission.
If the interaction is not very strong ($\lambda < 6\times 10^4$) then exchange effects are
substantial and affect tunneling. The developed WMD method allowed us to analyze some
interesting features of tunneling and proved to be a powerful tool for study of nonstationary
quantum processes. Of course, the numerical solution of Schr\"odinger equation would
not be too difficult for the problem under consideration, but in this work we just demonstrate
the method, regarding relatively simple system of two identical particles in the double-well
potential. We made only the first step in the development of WMD for the many-body problems and
we intend to report the results of investigation of many-particles system in the next paper.

\begin{acknowledgments}
The work was supported by RFBR, INTAS and Ministry of Sciences.
\end{acknowledgments}
\appendix

\end{document}